\documentclass[12pt]{spieman}  % 12pt font required by SPIE;
\usepackage{amsmath,amsfonts,amssymb}
\usepackage{graphicx}
\usepackage{setspace}
\usepackage{tocloft}
\usepackage{siunitx}
\usepackage{url}
\usepackage{tablefootnote}

\title{Warm Spitzer IRAC Photometry: dependencies on observing mode and exposure time}

\author[a]{Jessica E. Krick}
\author[a]{Patrick J. Lowrance}
\author[a]{Sean Carey}
\author[a]{Jason Surace}
\author[a]{Carl J.\ Grillmair}
\author[a]{Seppo Laine}
\author[a]{Schuyler D.~Van Dyk}
\author[a]{James G. Ingalls}
\author[b]{Matthew L.\ N.\ Ashby}
\author[b]{S.P. Willner}

\affil[a]{IPAC, MC 330-6, Caltech, 1200E. California Blvd. Pasadena, CA 91125}
\affil[b]{Center for Astrophysics $|$ Harvard \& Smithsonian, 60 Garden St.\ MS-66, Cambridge MA 02138}

\cftpagenumbersoff{figure}
\cftpagenumbersoff{table}

\begin{document} 
\maketitle

\begin{abstract}
We investigate differences in {\sl Spitzer}/IRAC 3.6 and 4.5\,$\mu$m photometry that depend on observing strategy.   Using archival calibration data we perform an in-depth examination of the measured flux densities (``fluxes'') of ten calibration stars, observed with all the possible observing strategies. We then quantify differences in the measured fluxes as a function of 1) array mode (full or subarray), 2) exposure time, and 3) dithering versus staring observations.  We find that the median fluxes measured for sources observed using the full array are 1.6\% and 1\% lower than those observed with the subarray at [3.6] and [4.5], respectively.  Additionally, we found a dependence on the exposure time such that for [3.6] observations the long frame times are measured to be lower than the short frame times by a median value of 3.4\% in full array and 2.9\% in subarray.  For [4.5] observations the longer frame times are 0.6\% and 1.5\% in full and subarray respectively. These very small variations will likely only affect science users who require high-precision photometry from multiple different observing modes.  We find no statistically significant difference for fluxes obtained with dithered and staring-modes.  When considering all stars in the sample, the fractional well depth of the pixel is correlated with the different observed fluxes.  We speculate the cause to be a small non-linearity in the pixels at the lowest well depths where deviations from linearity were previously assumed to be negligible.

\end{abstract}

% Include a list of up to six keywords after the abstract

% Include email contact information for corresponding author
{\noindent \footnotesize\textbf{*}Jessica Krick,  \linkable{jkrick@caltech.edu} }

\begin{spacing}{2}   % use double spacing for rest of manuscript

\section{Introduction}
\label{sect:intro}  % \label{} allows reference to this section
The Infrared Array Camera (IRAC) \cite{2004ApJS..154...10F} was operational on the {\it Spitzer} Space Telescope \cite{2004ApJS..154....1W} from 2003 - 2020 with four broad mid-infrared bands with response covering 3.15 - 9.25 $\mu$m (at 3.6, 4.5, 5.8, and  8.0 $\mu$m; also denoted channels 1 -- 4, respectively).  These filter names are labels and not the actual effective wavelengths (for more detailed information on filter transmission see Ref.~\citenum{2008PASP..120.1233H}).  After depletion of the cryogens, from mid 2009 until January 2020, only [3.6] and [4.5] were available for observations.

In order to efficiently calibrate the telescope, a discrete set of observing templates was enabled for IRAC observers.  Available options to define the templates included array mode, frame time, and dithering strategy.  The array mode could be specified either as full array (meaning, the entire IRAC 256$\times$256 array of 1.2\,arcsecond pixels was read out after each exposure) or subarray, in which case 64 consecutive 32$\times$32 pixel images were taken at a higher readout rate without moving the telescope (0.01~s  single readout for subarray vs. 0.2~s for full array).  A number of preset frame times were available; those and their corresponding exposure times are listed in Table~\ref{tab:frametimes}.  Frame time is a measure of time elapsed between successive array resets.  The exposure time (effective integration times)  is the time elapsed between the first pedestal and the first signal read, not including resets or multiple reads.  High dynamic-range exposures, in which paired short and long exposures were acquired sequentially within a single full-array observing template, were also available to IRAC observers, but did not add diversity to the range of possible observations.  Lastly, observers were able to choose from a discrete set of dither patterns, make their own mapping strategy, or employ a staring-mode.  Dithering (small position changes to reduce noise) and mapping (position changes of order the size of the array to cover an area greater than the size of the array) were both possible in the full array mode.  Dithering was available in the subarray mode between 64-frame sets, but not within the sets.  Observations which do not move positions between frames are known as ``staring-mode.''  Staring-mode was most often used for high-precision time-series observations of brown dwarfs and exoplanets.

%For Table\ref{tab:frametimes}, note that we also tested a 400s frame time for the warm mission but didn't use either the 200 or 400s as the benefits in sensitivity were outweighed by the increased number of cosmic ray affected pixels and the lack of dither diversity in most observations.

\begin{table}[ht]
\caption{Available IRAC Frame Times} 
\label{tab:frametimes}
\begin{center}       
\begin{tabular}{|l|l| l|} %% this creates two columns
%% |l|l| to left justify each column entry
%% |c|c| to center each column entry
%% use of \rule[]{}{} below opens up each row
\hline
\rule[-1ex]{0pt}{3.5ex}  Frame Time & Exposure Time & Observing Mode \\
\rule[-1ex]{0pt}{3.5ex}  (seconds) & (seconds) &  \\

\hline\hline
\rule[-1ex]{0pt}{3.5ex}  0.02 & 0.01  & subarray\\
\hline
\rule[-1ex]{0pt}{3.5ex}  0.1  & 0.08 & subarray  \\
\hline
\rule[-1ex]{0pt}{3.5ex}  0.4 & 0.36 & subarray   \\
\hline
\rule[-1ex]{0pt}{3.5ex}  2 & 1.92 & subarray  \\
\hline \hline
\rule[-1ex]{0pt}{3.5ex}  0.4 & 0.2 & full array  \\
\hline
\rule[-1ex]{0pt}{3.5ex}  2 & 1.2 & full array \\
\hline
\rule[-1ex]{0pt}{3.5ex}  6 & 4.4 & full array\\
\hline
\rule[-1ex]{0pt}{3.5ex}  12 & 10.4 & full array \\
\hline
\rule[-1ex]{0pt}{3.5ex}  30 & (23.6, 26.8) \tablefootnote{ Entries with two exposure times are for [3.6] and [4.5] respectively.} & full array \\
\hline
\rule[-1ex]{0pt}{3.5ex}  100 & (93.6, 96.8) & full array \\
\hline 
%\rule[-1ex]{0pt}{3.5ex}  200 & (187.2, 193.6) & full array \\
%\hline 

\end{tabular}

\end{center}
\end{table}

The rich {\sl Spitzer}/IRAC archive contains many observations of the same target taken with different array modes, frame times, or dithering strategies.  This is often the case for serendipitous observations but also for observations designed intentionally in different modes.  Possible science cases for this include, but are not limited to, 1) initial dithered observations to find proper motions of brown dwarfs followed by staring-mode data to refine the characterization of their atmospheres, 2) dithered debris disk searches looking for IR excess followed by a staring-mode study of variations in specific debris disks, 3) archival dithered observations followed by targeted subarray staring observations for confirmation, or 4) observations that had either a too low signal-to- noise ratio (SNR) or were saturated at one exposure time were followed up later in the mission using a different exposure time.  These science cases therefore require combining photometry from different modes to arrive at scientific conclusions.

In this work we examine how IRAC photometry depends on observing strategy.  We emphasize that these are very small measured variations (a few percent at most) and so will likely only affect science users who require high-precision photometry from multiple different observing modes.  A full set of calibration observations to test for variations in measured fluxes among observing modes were only taken during the warm IRAC mission. We consequently do not discuss data taken during the cryogenic mission (which ended in 2009), or in the [5.8] or [8.0] channels.

In Section \ref{sec:data} we describe the archival data used for this project.  Section~\ref{sec:photometry} covers our methods for reducing the data, carrying out photometry, and applying photometric corrections.  Section~\ref{sec:results} discusses the different potential correlations and photometry effects.  We make concluding remarks in Section~\ref{sec:conclusions}.

\section{The Data}
\label{sec:data}

\subsection{Observations}
\label{sec:observations}
For this work we used the calibration observations taken for {\sl Spitzer} Program ID (PID) 1336 and PID 1367.  Specifically, we observed a set of ten stars with varying exposure times, in both subarray and in full array, and in staring and dithering modes.  Almost always, staring-mode observations were taken on the same pixel ("the sweet spot"), as that pixel was the best characterized pixel on the array.  Table~\ref{tab:stars} lists the stars.  We searched the archive for additional data usable for this analysis but did not find anything suitable, combining non-variable stars with observations in all available modes and having a sufficient number of images to achieve statistical significance.

\begin{figure}
\begin{center}
\begin{tabular}{c}
\includegraphics[height=9cm]{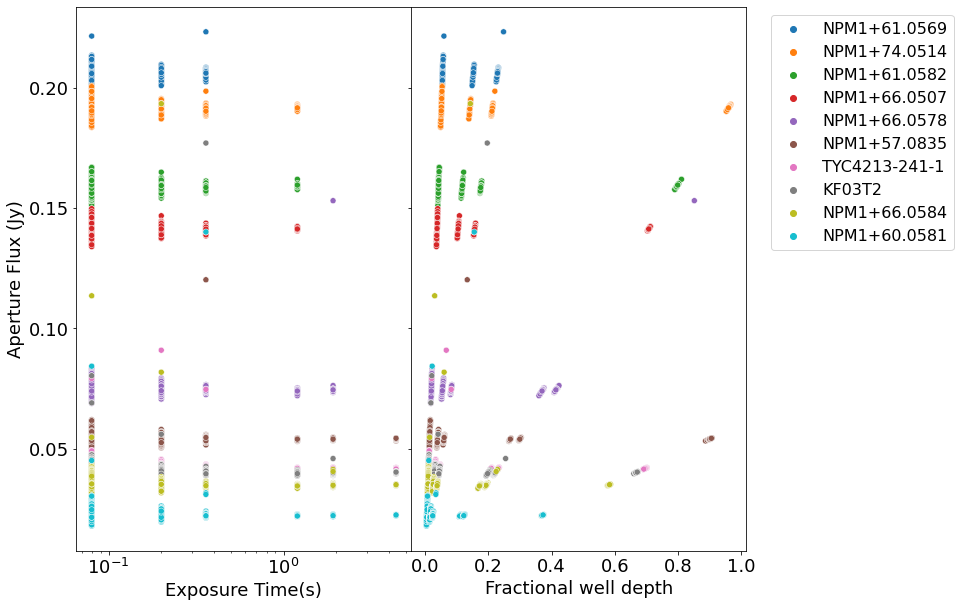}
\end{tabular}
\end{center}
\caption 
{ \label{fig:flux_DN}
Description of the dataset used in this work for the [4.5] channel, showing the aperture flux as a function of the exposure time on the left panel and aperture flux as a function of well depth on the right panel.  The [3.6] channel has a similar observation set but for brevity is not shown. Points are color coded by star names, and symbol shapes denote full (filled circles) or subarray (crosses).  Each data point represents one observation, and data are not binned, but there are overlapping data points. } 
\end{figure}

Figure~\ref{fig:flux_DN} shows a visualization of this dataset.  We plot the exposure time(s) vs.\ aperture flux for just the [4.5] channel.  Similar observations were made for [3.6]. All stars were observed with multiple exposure times.  Not all of the ten stars could be observed at all exposure times due to SNR and saturation concerns.  This plot illustrates how the range of stellar brightnesses and frame times in the
sample filled the available phase space.
%This plot shows that we attempted to fill the available phase space by varying star brightness within the sample and by varying frame time.  

We considered how close to saturation(well depth) a star is  as a means of interpreting our results.  Looking at our sample as a whole, the range of possible well depths is not well sampled;  having a median fractional well depth of 0.04 and 0.02 at [3.6] and [4.5], where a fractional well depth of 1.0 indicates saturation. The right panel of Figure~\ref{fig:flux_DN} shows well depths of the sample. The median well depths of our sample correspond to a SNR of 39 and 24 at [3.6] and [4.5], respectively. This dataset was not designed with well depth in mind, and was instead designed to find stars which would sample the available exposure time parameter space.

Although available, we reject the 0.02~s photometry because of its large scatter. We have also rejected any data where the well depth is greater than the listed saturation limit in the IRAC Instrument Handbook.\footnote{\label{handbook}\url{https://irsa.ipac.caltech.edu/data/SPITZER/docs/irac/iracinstrumenthandbook/}}  The saturation limit results in the rejection of around 250 photometry points at [4.5].   All observations used in this work have SNRs of six or greater.

For a subset of three stars, we made a larger number of observations in all possible modes  with 2~s frame times in both channels.  This allowed us to build up statistically meaningful samples where we hold the frame time constant while varying the other observing parameters. The three stars are NPM1+66.0584, NPM1+66.0578, and KF03T2.  

The 2~s frame time is the closest set of IRAC frame times available for holding exposure time constant between the full and subarray. Unfortunately, the 2~s subarray and 2~s full-array frame times do not actually correspond to the same exposure times; 1.92~s and 1.2~s, respectively.  Similarly, frame times of 0.4~s in full and subarray also have different exposure times, so that frame time was not a good alternative.   This difference in the exposure times for the same frame time comes from a different number of Fowler numbers (N), wait ticks, and readout times for the different modes as listed in the IRAC Instrument Handbook. Briefly, Fowler sampling is a way of observing by taking N non-destructive reads at the beginning of the observation and another N non-destructive reads at the end of the observation, the difference of which is the flux measured per pixel.

All of the stars chosen for this work were vetted as potential calibration stars for IRAC.  Seven of the ten stars are published as primary or secondary calibrators for IRAC known to not exhibit flux variability in the IRAC bands \cite{2005PASP..117..978R}. The three stars not included in that reference are NPM1+74.0514, NPM1+57.0835, and NPM1+66.0584.  Specifically NPM1+66.0584 is one of the three stars in the subset taken with more observations in 2~s frame times.  We cannot use this dataset to determine both if the stars are time variable, and if they vary as a function of the other parameters studied herein.  While vetted,  because these have not specifically been published as calibration stars, we experimented with removing these three stars from our sample.  All plots look similar (albeit with larger scatter due to fewer data points), and conclusions remain the same if we remove those stars from the sample.  We therefore choose to keep these in the sample for the remainder of this work.

  Each of the [3.6] and [4.5] datasets include about 80,000 total individual observations. Specifically, for each channel we have roughly 65,000 subarray and 15,000 full array observations.

\begin{table}[ht]
\caption{The Calibration Star Sample} 
\label{tab:stars}
\begin{center}       
\begin{tabular}{|l l l l l l l|} %% this creates two columns
%% |l|l| to left justify each column entry
%% |c|c| to center each column entry
%% use of \rule[]{}{} below opens up each row
\hline
\rule[-1ex]{0pt}{3.5ex}  Name & R.A. & Dec. & Spec.  & [3.6] & [4.5] & Alternate \\
\rule[-1ex]{0pt}{3.5ex}  &  (h m s) &  (deg' '') & Type & (mJy) & (mJy) & Name\\

\hline\hline
\rule[-1ex]{0pt}{3.5ex} NPM1+61.0569 & 17 23 25.9 & +61 12 40.7 &K0.5 III & 363	& 208 & \\
\hline
\rule[-1ex]{0pt}{3.5ex} NPM1+74.0514 &	19 02 53.5& 	+74 14 43.6	&K0.5 III&	323&	194&	BD+74 804  \\
\hline
\rule[-1ex]{0pt}{3.5ex}  NPM1+61.0582 &	17 36 55.6&	+61 40 58.1	&K1 III &	274&	162 & \\
\hline
\rule[-1ex]{0pt}{3.5ex}  NPM1+66.0507 &	17 31 22.1&	+66 46 35.3&	K2 III&	250&	145& \\
\hline 
\rule[-1ex]{0pt}{3.5ex}  NPM1+66.0578 &	19 25 32.2 &	+66 47 38&	K1 III &	127&	75 & \\
\hline
\rule[-1ex]{0pt}{3.5ex}  NPM1+57.0835	&18 04 12.6	&+57 42 18.6&	&	95&	54& \\
\hline
\rule[-1ex]{0pt}{3.5ex}  TYC4213-241-1&	18 03 45.53	&+66 56 03.7&	K1 III &	74	&42& KF01T3\\
\hline
\rule[-1ex]{0pt}{3.5ex}  KF03T2 &	17 57 51.4&	+66 31 03.0 &	K1.5 III&	70&	40 &\\
\hline
\rule[-1ex]{0pt}{3.5ex}  NPM1+66.0584&	19 36 07.39&	+66 21 54.2&	F0	&56	&35	&BD+66 1222 \\
\hline
\rule[-1ex]{0pt}{3.5ex}  NPM1+60.0581 &	17 24 52.3&	+60 25 50.8	&A1 V &	37&	23&	BD+60 1753 \\
\hline

\end{tabular}

\end{center}
\end{table}

\subsection{Photometry}
\label{sec:photometry}

We briefly describe our pipeline here, emphasizing where it is differs from previous work.  For an overview of the {\sl Spitzer}/IRAC absolute photometric calibration, see Refs.~\citenum{2005PASP..117..978R, 2012SPIE.8442E..1ZC}.  Specifics of the photometry pipeline used here are included in Ref.~\citenum{2021AJ....161..177K}. We used the basic calibrated data (BCD) FITS files (suffix bcd.fits) from the Spitzer Heritage Archive (SHA).  These BCD files have had a dark correction, flat-field correction, linearity correction, and conversion to flux units already applied.

One somewhat novel aspect of this work is that we apply a different dark correction to staring-mode data than we do to dithered observations.  Staring-mode data are processed in the standard pipeline in the same manner as dithered data, including using a dark image which was made by dithering.  The delay time between frames affects the bias level in the frame (including in the dark); this is known as the ``first-frame effect.''  Therefore, dithered observations will have different bias levels and patterns than staring-mode observations since staring-mode observations have shorter delay times between consecutive frames, although that pattern is constant if the delay time is constant.  This effect adds both noise and systematics to the photometry.  For this reason, dithered darks are inappropriate for staring-mode science frames when precision photometry is required.   

We used PID 1345 to make our own staring-mode dark suite for all subarray frame times. We began by removing the dithered dark correction from the PID 1345 data.  Because the dark correction is not the last correction made to the BCD files, care was taken to first back out the other corrections, apply the staring-mode dark, then  re-apply the other pipeline corrections. We then created a median image for each exposure time (0.02, 0.1, 0.4, 2.0 s) and used this median frame as the staring-mode dark.  Applying this new staring-mode dark to the staring-mode data has a measurable impact on derived fluxes. We recommend that anyone doing precision photometry with staring-mode data use a staring-mode dark instead of the pipeline-provided, dithered dark.  While the IRAC pipeline will not include these starting-mode darks, code is available for users to to change which darks are used in a BCD frame on the contributed-code section of the {\it Spitzer} IRAC website. \footnote{\url{https://irsa.ipac.caltech.edu/data/SPITZER/docs/dataanalysistools/tools/contributed/irac/change_dark_calibrate/}}

To measure flux, we use our appropriately dark-corrected BCD exposures and make the following corrections in order.  We first convert images into units of electrons to enable a statistical calculation of uncertainties.  Second, we use a center-of-light method to find stellar centroids. \footnote{\url{https://irsa.ipac.caltech.edu/data/SPITZER/docs/irac/calibrationfiles/pixelphase/box_centroider.pro}}  Third, we do aperture photometry with a three-pixel radius aperture and (3-7) pixel background annulus.  The small aperture size is chosen to reduce noise and the number of cosmic rays in the aperture.  Fourth, we make a correction for pixel-phase using pixel\_phase\_correct\_gauss.pro. \footnote{\url{https://irsa.ipac.caltech.edu/data/SPITZER/docs/irac/warmfeatures/pixel_phase_correct_gauss.pro}}  The pixel-phase correction accounts for gain changes  as  a  function  of the position  within  a pixel, coupled with the undersampling of a point source by IRAC.  Fifth, we make a correction for array location. The array location-dependent correction takes into account the variation in system response of the instrument across the field of view, which is primarily  due  to  the  change  in  the angle  of  incidence  of light through the bandpass filter as a function of position on the array. Lastly, we discard the first frame of every subarray FITS file and of every full array AOR.  These frames are affected by the first-frame effect discussed above and are likely to have measured fluxes that differ from those of subsequent images. We do not apply an aperture correction since the same aperture is used for all photometry regardless of observing mode.

To compare photometry for all stars on the same plots, we normalize the stars to the same absolute level by dividing all photometry by the median stellar fluxes.  The distributions of fluxes per star are somewhat skewed, so a mean does not capture the peak of the distribution. Having skewed distributions causes the mean levels in the subsequent plots to differ from unity.

\section{Results and Discussion}
\label{sec:results}
This section describes how the measured IRAC photometry differs as a function of observing mode and exposure time. Sec.~\ref{sec:disentangle} examines the effects of observing mode and exposure time together. Finally, we consider the impact of staring and dithering on IRAC photometry.

We choose to present our results mainly with box plots to show the distributions per star as a function of both array mode (full or subarray) and exposure time for both channels, discussed individually below in Sections~\ref{sec:readmode}~and~\ref{sec:exposure time}.  A box plot shows the median of the distribution as the solid line in the middle of the box. The box top and bottom indicate the top and bottom quartiles and the caps at the end of the lines show the maximum and minimum values in the distributions. These box plots include the entire dataset (all stars, all exposure times, full and subarray, staring and dithering).  On average, each box contains a few thousand data points. Generally the boxes with smaller quartile ranges and max/min values have fewer data points. The plots are color-coded by star for each of the ten stars.  Stars are listed in order from the brightest (NPM1+61.0569) to the faintest (NPM1+60.0581).  Color-coding is consistent for all box plots and labeled in Figure~\ref{fig:readmod_ch1}.

\subsection{Array Mode - Full vs. Subarray }
\label{sec:readmode}

\begin{figure}
\begin{center}
%\begin{tabular}{c}
\includegraphics[width = 0.75\linewidth]{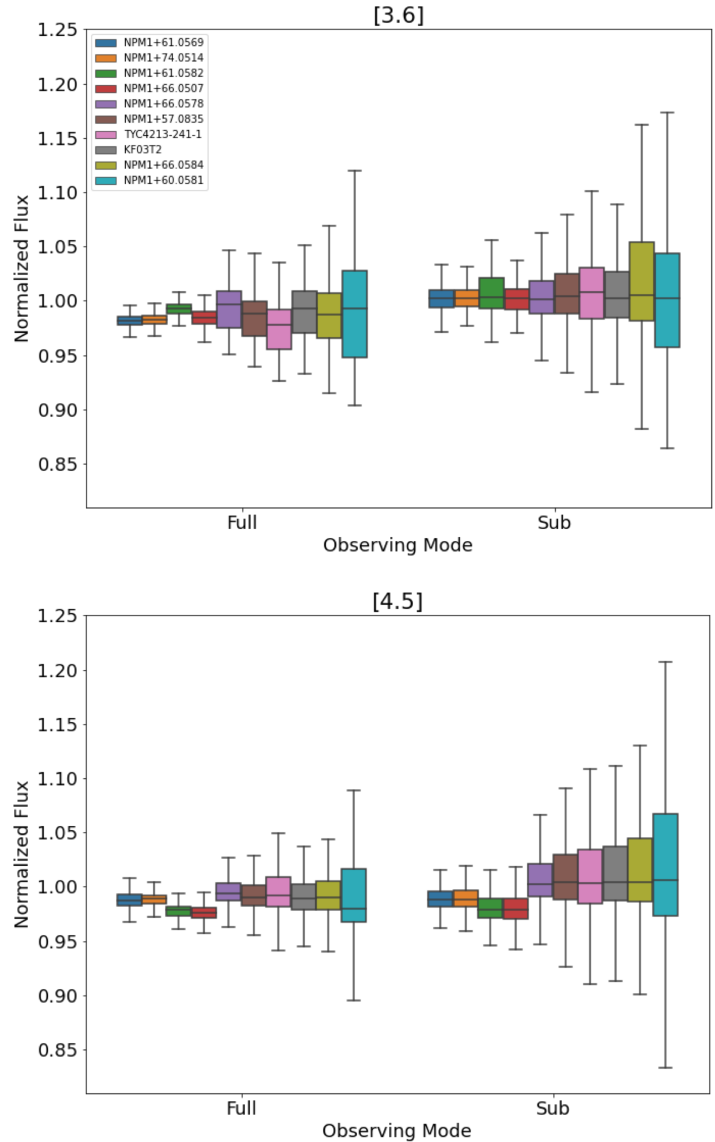}
%\includegraphics[width = 0.75\linewidth]{areadmode_box_ch1_noff_staredark.png}
%\includegraphics[width = 0.75\linewidth]{areadmode_box_ch2_noff_staredark.png}
%\end{tabular}
\end{center}
\caption 
{ \label{fig:readmod_ch1}
Box plot of the normalized fluxes of all the stars divided into full and subarray observations for [3.6] on the top and [4.5] on the bottom. Stars are normalized by dividing by the median flux.  The solid line in the middle of the boxes shows a median value, while the top and bottom of the box are quartiles, and the endcaps indicate the full range of the data.  Stars are color coded according to the legend and listed in order of brightness from brightest to faintest. On average, the measured fluxes in full-array observations are smaller than the measured fluxes in the subarray observations. } 
\end{figure}

Figures~\ref{fig:readmod_ch1} shows the [3.6] and [4.5] photometry (top and bottom respectively) of the stars in Table~\ref{tab:stars} as a function of array mode, revealing that the measured fluxes appear on average to be lower for the full-array observations than the subarray observations of the same stars. We attempt to confirm this statistically by using an Anderson two-sample test, per star, to see if the distributions for full array and subarray are drawn from the same population. This statistical method considers the vertical distance between the two cumulative distributions. For all ten stars we can say that the full array and subarray data are not drawn from the same distribution at the 25\% significance level (maximum possible significance).  Thus, the difference between the fluxes measured in full array and subarray modes are statistically significant. Using all ten stars, the median difference between the full and subarray flux measurements is 1.6\% at [3.6] and 1.0\% at [4.5].

\subsection{Frame Time}
\label{sec:exposure time}

\begin{figure}
\begin{center}
%\begin{tabular}{c}
\includegraphics[width = 0.75\linewidth]{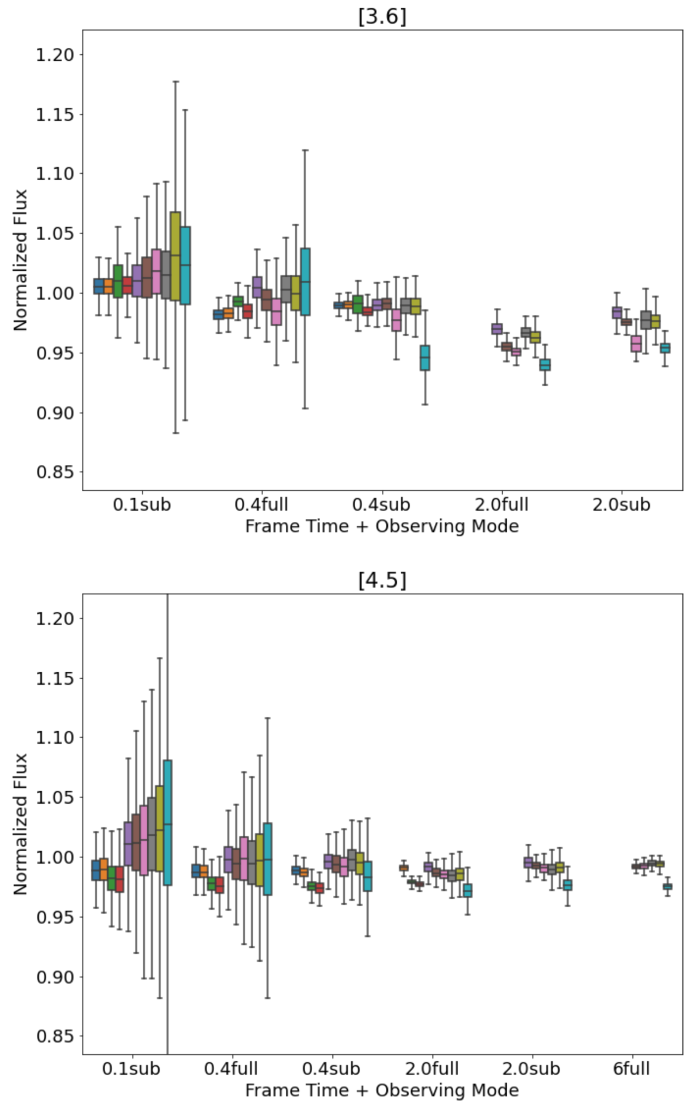}
%\includegraphics[width = %0.75\linewidth]{exptime_box_ch1_noff_staredark.png}
%\includegraphics[width = %0.75\linewidth]{exptime_box_ch2_noff_staredark.png}
%\end{tabular}
\end{center}
\caption 
{ \label{fig:exptime_ch1}
Box plot of the distributions of normalized fluxes of all the stars as a function of exposure time at [3.6] on the top and [4.5] on the bottom. Frame time labels also indicate full or subarray.  Color coding is the same as for Figure~\ref{fig:readmod_ch1}.  On average, fluxes measured at longer frame times are lower than those measured at shorter frame times.} 
\end{figure}

\begin{figure}
\begin{center}
\begin{tabular}{c}
\includegraphics[height=8cm]{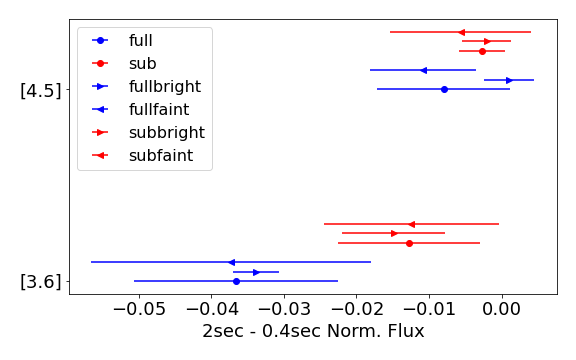}
\end{tabular}
\end{center}
\caption 
{ \label{fig:frametime_effect}
Exposure-time effect for both channels divided into bright and faint stars to look for an effect with star brightness.  [4.5] is shown in the top half of the diagram, while [3.6] plotted in the lower half.  Blue circles show the median difference for all the stars observed in the full array between the 2~s and 0.4~s frame time normalized fluxes, including one sigma error bars.  Blue right and left triangles show the median values for the bright and faint populations of stars.  The same quantities are plotted for subarray data in red.} 
\end{figure}

Figure~\ref{fig:exptime_ch1} shows the distributions of normalized calibrator star fluxes as a function of Frame time.  [4.5] includes more data on the six-second full-array frame time, so the bottom plot includes that frame time whereas the [3.6] plot does not. The median normalized flux of all the stars changes as a function of frame time such that the measured fluxes are larger at lower frame times.  This is most strongly evident in [3.6].  Figure \ref{fig:frametime_effect} shows a measure of the strength of this effect as the median difference between the 2~s and 0.4~s frame time fluxes for all the stars, per channel, per observing mode.

As in Section~\ref{sec:readmode}, we used the Anderson two-sample test to compare the full array 0.4~s observations with the full array 2~s observations.  For both channels, we find the difference between the normalized flux distributions at the two frame times to be statistically significant.  The same is true for fluxes from the subarray in 0.4~s and 2~s frame time observations. 

We tested whether the trend seen in flux as a function of exposure time depends on stellar brightness such that the brighter or the fainter stars would be more or less likely to show this effect.  We do this by calculating the difference between the 2~s and the 0.4~s normalized fluxes.  A difference between the fluxes implies that different fluxes are measured using different exposure times.  Because we know there is an effect with array mode, we divide the sample into full and subarray data.  Finally, looking for a trend in the stellar brightness, we divided the sample into bright and faint stars, and re-calculated the median values.  Figure~\ref{fig:frametime_effect} shows the results, where the circular points are the median values of the difference in fluxes between the 2~s and 0.4~s frame-times for both the full array only data (blue) and the subarray only data (red). The bright and faint subsets are shown with the appropriately colored triangles. The differences between the bright and faint samples are within one sigma of each other. Here sigma is calculated as the standard deviation between stars, which means that each bin contains only a few stars (and not thousands of data points).  Our sample size is not large enough to make conclusions based on star brightness.

\subsection{Disentangling Exposure Time and Array Mode Effects}
\label{sec:disentangle}
We examine the possibility of disentangling the effects of exposure time and array mode. Instead of dividing the dataset by star, here we consider all stellar photometry as a single dataset, and divide the dataset into four categories depending on the combination of array mode (full array vs.\ subarray) and exposure time (short vs.\ long). Each distribution has between 3000 and 50000 stars (subarray exposure times have lots more frames than full array). Exact exposure times cannot be compared in this analysis because they differ significantly for full and subarray modes (see Table~\ref{tab:frametimes}).  For that reason, the division between the short and long exposure times is set at 0.3~s, to construct statistically significant samples. 

\begin{figure}
\begin{center}
%\begin{tabular}{c}
\includegraphics[width = 0.75\linewidth]{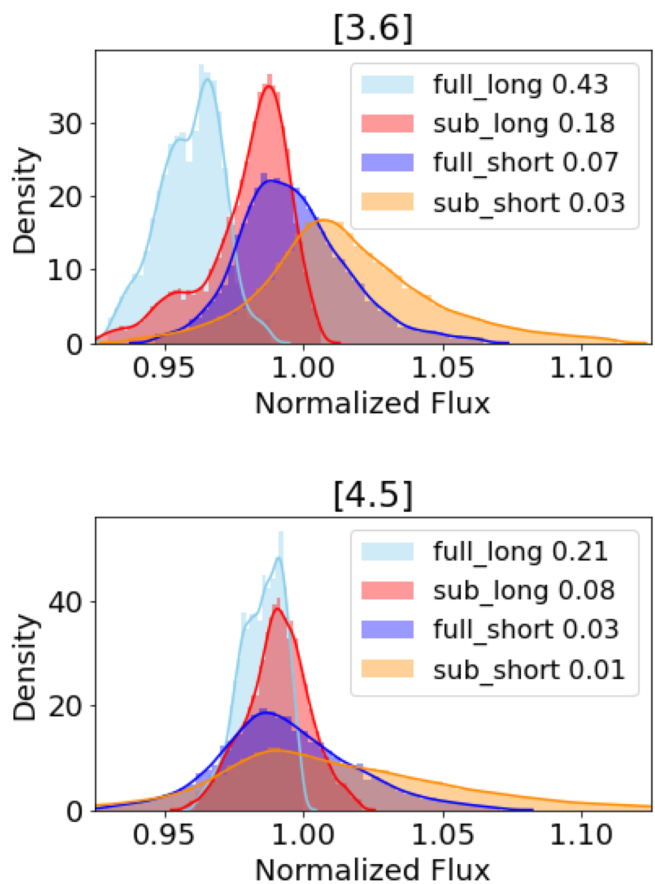}
%\includegraphics[width = 0.75\linewidth]{corrflux_dist_short_long_ch1.png}
%\includegraphics[width = 0.75\linewidth]{corrflux_dist_short_long_ch2.png}
%\end{tabular}
\end{center}
\caption 
{ \label{fig:corrflux_dist_ch1}
Distribution of normalized star fluxes for [3.6] on the top and [4.5] on the bottom.  The legend lists the different observing modes (full array vs subarray and long exposure times vs. short exposure times) as well as the mean well depth per distribution.  Solid dark lines show the kernel density estimation (KDE) overplotted on the histograms. We see a difference in the absolute photometry of stars taken in  available observing modes and exposure times shown here in the colored distributions.  } 
\end{figure}

Figure~\ref{fig:corrflux_dist_ch1} reveals differences between the overall distributions of photometry taken in full and subarray mode while accounting for exposure time, with enough data points to accurately reflect the shapes of the distributions.  Longer exposure times are in red and light blue; shorter exposure times are in orange and dark blue.  Especially at [3.6], an effect is apparent with both exposure time and observing mode. For the stars which have full-array observations in both short and long exposure times, the median difference in the flux is  3.4\% at [3.6] (for six stars) and  0.6\% at [4.5] (for ten stars). For subarray we find the median difference in flux between short and long exposure times to be 2.9\% at [3.6] and 1.5\% at [4.5]. 

%We do an experiment with [3.6] of statistically correcting the data for the full vs. subarray difference as listed above to see if the exposure time effect goes away. We correct the ch1 full array data by 1.6\% to bring it in line with the ch1 sub array data (shifting the green and red histograms). The difference between short and long exposure times remains.  This indicates that there are two different effects at work here.

\subsection{Well Depth}
\label{sec:well_depth}
The legend to Figure~\ref{fig:corrflux_dist_ch1}  lists median well depths for each of the distributions.  Well depth is one physical feature which correlates with the difference between short and long and full and subarray exposures.  We do expect that longer exposure times on the same set of stars will have larger fractional well depths, so it makes sense that the full long and sub long distributions have the larger median well depths in both channels.  Also, subarray mode has the possibility of shorter frame times than full array, so we would expect that subarray would have lower well depths than full array.  Both channels show this behavior.

\begin{figure}
\begin{center}
%\begin{tabular}{c}
\includegraphics[width = 0.75\linewidth]{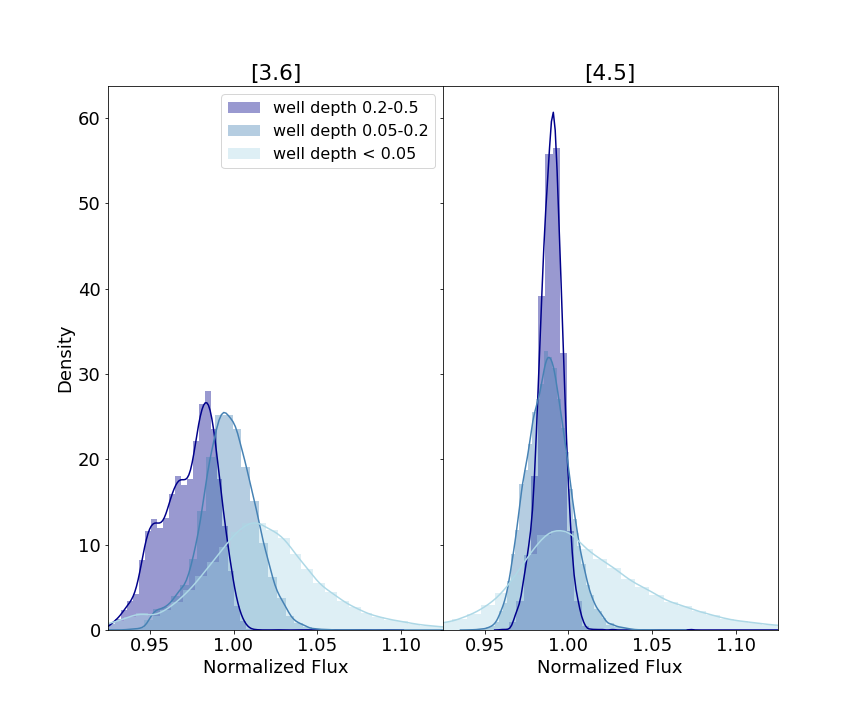}

%\end{tabular}
\end{center}
\caption 
{ \label{fig:welldepth_normflux}
Distributions of fluxes in three fractional well depth bins for [3.6] on the left and [4.5] on the right. Colors for the distributions get darker as the fractional well depth increases.} 
\end{figure}

Figure~\ref{fig:welldepth_normflux} shows the distributions of normalized fluxes for the entire sample of stars in [3.6] and [4.5] divided into three well-depth bins.   These bins do not have the same number of photometry points in them as we have  many more low well-depth observations than intermediate well-depth observations.  Because of this change in the number of data points per bin, we do not quote values for the peak of each bin, but rather address the trends revealed by the data in hand [3.6] shows a clear trend such that observations at intermediate well depths have lower normalized fluxes.  [4.5] shows little, if any trend, consistent with trends seen in the observing mode and exposure time plots.  While there are sources at higher well depth than 50\% full well, their distribution overlaps those of the 20-50\% bin. We conclude that only fractional well-depths less than a few tens of percent are affected by this well-depth effect.  Consideration of well-depth does not help to disentangle the effects we have seen with observing mode and exposure time, but it is a clue that potentially the source of some of what we are seeing is non-linearity in the low well-depth regime.

\subsection{Dithering vs. Staring-mode}
\label{sec:stare_dither}
 We do not detect a difference in photometry between the dithered and staring-mode observations.  For three stars we have observations in all the modes for the 2~s exposure times (sub staring, sub dithering, full staring, full dithering). Recall that subarray observations taken with 2~s frame-time have 1.92~s exposure times and 2~s frame time full-array observations have 1.2~s effective exposure time, so the 2~s exposure time subarray and full array observations are not equivalent, but they are the closest to equivalent that exist in the archive.

\begin{figure}
\begin{center}
\begin{tabular}{c}
\includegraphics[height=8cm]{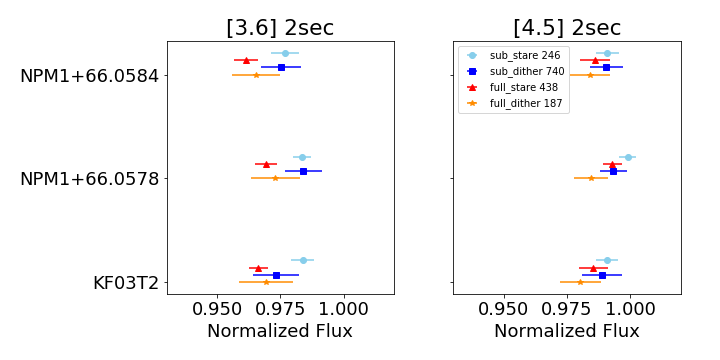}
\end{tabular}
\end{center}
\caption 
{ \label{fig:cleveland_dot}
Cleveland dot plot comparing normalized fluxes from different observing modes and dithering vs. staring for three stars listed on the y-axis.  [3.6] is on the left, [4.5] is on the right.  The legend is for both plots, and lists the number of data points for each mode.  } 
\end{figure}

Figure~\ref{fig:cleveland_dot} shows a Cleveland dot plot of the median and standard deviation of the normalized fluxes for each mode, for each of the three stars in this sample. The legend also lists the number of data points per mode.  The full and subarray measurements differ as before( see Section~\ref{sec:readmode}). The staring-mode photometry is consistent with the dithered mode photometry within one sigma in both channels. 

%The photometry separately measured in the staring mode and subarray observations is consistent within one sigma, in both IRAC channels. 

\subsection{Multiple Regression}
Finding a relation among the observing parameters (array mode, exposure time, staring/dithering) may help both to a) understand the effects found in this paper and b) correct for them. We therefore use the statistical technique of multiple regression to search for any relations.   Specifically, we used Multiple Linear Regression, i.e., multiple independent variables, to predict the value of the dependent variable.  To correct IRAC photometry for these effects, we have tried ordinary least squares (OLS) as a multiple linear-regression technique.  We used array mode, exposure time, and stare/dither as independent variables, and flux as the dependent variable.  We used two different modules in Python for this work, statsmodels\cite{seabold2010statsmodels} and sklearn\cite{scikit-learn}.  We are unable to find successful models.  The $R^2$ goodness of fit is 0.037, when "good" models should have values close to 1.0.  The failure of OLS in this situation could imply that the relationship between the independent and dependent variables is nonlinear.

%\begin{figure}
%\begin{center}
%\begin{tabular}{c}
%\includegraphics[height =6cm]{KF03T2_ch1.png}
%\end{tabular}
%\end{center}
%\caption 
%{ \label{fig:KF03T2_ch1}
%Normalized flux of calibration star KF03T2 at [3.6] as a function of time since the first observation in each mode. Points are color coded based on observing mode.  } 
%\end{figure} 

%\begin{figure}
%\begin{center}
%\begin{tabular}{c}
%\includegraphics[height =6cm]{KF03T2_ch2.png}
%\end{tabular}
%\end{center}
%\caption 
%{ \label{fig:KF03T2_ch2}
%Same as figure~\ref{fig:KF03T2_ch1} except for [4.5]  } 
%\end{figure} 

\section{Conclusions}
\label{sec:conclusions}
We document exposure time, array mode (full vs. subarray), and fractional well depth flux dependence in IRAC photometry. The full array has median fluxes higher than the subarray  by 1.6\% at [3.6] and 1.0\% at [4.5].  Dividing the sample further, the long exposures have a lower median flux than shorter exposures.  The difference at full array is 3.4\% at [3.6] and 0.6\% at [4.5]. For subarray we find the median difference in flux is 2.9\% at [3.6] and 1.5\% at [4.5]. These noted effects are only relevant for a small fraction of IRAC high precision users with data in multiple modes who should include these values in their uncertainty calculations. We posit two potential causes of the noted low-level differences in photometry.

Overall, the normalized fluxes decrease as well depth increases up to a few tens of percent full well.  Well depth is correlated with the differences in the flux distributions between full and subarray and long and short exposures.  While these correlations with array mode and exposure time are expected, it potentially indicates the presence of non-linearities at the low well depths sampled in this work.  

Linearity corrections were made for warm IRAC to correct for a known effect where an increase in incoming photons does not correspond to an increase in counts (or data numbers DN).  This occurs because filling the well decreases the potential, which in turn results in a less responsive system.  The linearity correction for the IRAC InSb arrays inherently assumes no linearity correction at low well depths. Therefore, time-intensive observations were not made during the mission to include extremely low well-depth observations. Instead, calibration observations focused on the moderate ($>$ 20\%) to high well depths where the linearity deviated most significantly from a straight line correlation between photons and DN (see IRAC data handbook for a description of the derivation of the warm linearity correction. \footnote{\url{https://irsa.ipac.caltech.edu/data/SPITZER/docs/irac/iracinstrumenthandbook/}})  One explanation for why this low count linearity effect could exist at [3.6], but not [4.5] is that the applied biases are different between [3.6] and [4.5] implying that the electric potentials are different, which could explain the stronger effect at [3.6] than [4.5].  

Understanding the root cause of non-linearities is beyond the scope of this paper.  Many complicating issues are hiding under that designation including the 3D structure of a pixel, how the linearity interacts with Fowler sampling (specifically how to translate corrections derived at one set of Fowler sampling parameters to those in another, which would get worse as the integration times become comparable to the time spent reading the detector), the speed at which the electric fields are changing in these very short exposures compared to the timescales of the exposures or the readouts and persistent image trap filling.

Linearity can in some detectors depend on the flux of the source in the sense that it matters not only how many photons come in to the detector, but also the rate at which they fill the well.  We do not have enough observations to know if the effect we are seeing is dependent on the brightness of the stars.

A second possible explanation for the difference between full and subarray photometry is that the array is resetting faster between consecutive frames for the subarray.  A faster reset applies a stronger reverse bias on the array more frequently, which could affect the distribution of photoelectron traps, and therefore could have a low-level effect on photometry. We know that other similar effects ( e.g., "first-frame effect") are more significant at [3.6] than [4.5].\footnote{\url{https://irsa.ipac.caltech.edu/data/SPITZER/docs/files/spitzer/som12.2.pdf}}

No flux difference is apparent between the staring and dithering mode observations after correcting the staring-mode data by a staring-mode dark.  Anyone doing absolute photometry with staring-mode data should be using a staring-mode dark.  The full and subarray staring-mode datapoints considered here are all taken on the same array pixel (the sweet spot of the subarray), which is not true for the dithered positions.  Thus, residual pixel-phase uncertainties cannot be the cause of the measured flux differences between the full and subarray staring-modes, otherwise we would see this effect in this work when comparing staring and dithering modes.

We have no evidence for low level persistent images being the cause of the differences in photometry measured here.  We know that the persistent images are stronger at [3.6].  However, for [3.6] the same fraction of observations were taken at the sweet-spot pixel in both full and subarray images.  The sweet spot is the pixel at the center of the array where most observations are conducted in the subarray because it has been the best characterized for the pixel phase effect.  This means that the sweet spot pixel is more likely to have frequent low-level persistent images than other pixels on the array.  If a larger fraction of observations in the subarray had been taken at the sweet spot than the full array, we might have expected persistent images to be the culprit.  On the contrary, we see no evidence for this.

Our analysis implies that these differences in fluxes are systematic; they do not average out with more observations.  All full-array photometry will be different than all subarray photometry, no matter how many observations are taken in any given observing mode.

\subsection* {Acknowledgments}
We thank the anonymous referees for their time and care in providing very useful comments on this manuscript. 
This work is based in part on observations made with the {\it Spitzer} Space Telescope, which is operated by the Jet Propulsion Laboratory, California Institute of Technology under a contract with NASA. This research has made use of NASA's Astrophysics Data System, the NASA/IPAC Infrared Science Archive, which is operated by the Jet Propulsion Laboratory, California Institute of Technology, under contract with the National Aeronautics and Space Administration, and the SIMBAD database, operated at CDS, Strasbourg, France.  The acknowledgements were compiled using the Astronomy Acknowledgement Generator. 
%This unnumbered section is used to identify those who have aided the authors in understanding or accomplishing the work presented and to acknowledge sources of funding. 

%\subsection* {Code, Data, and Materials Availability} 
%As relevant, declare the availability of computer software code, data, and/or materials used in the research results reported in the manuscript. Provide specific access information or restrictions for code, data, and materials (i.e., links to repository access addresses, and/or guidance on commercial or public access). Note: reporting in this section is required for the \textit{Journal of Biomedical Optics} and \textit{Neurophotonics}. 

%%%%% References %%%%%

\bibliography{report}   % bibliography data in report.bib
\bibliographystyle{spiejour}   % makes bibtex use spiejour.bst

\listoffigures
\listoftables

\end{spacing}
\end{document}